\documentstyle[aps,preprint]{revtex}

\begin{document}

\title{Coherent and squeezed condensates in massless $\lambda\varphi^4$ theory}

\author{H.-P. Pavel, D. Blaschke and G. R\"opke}
\address{MPG Arbeitsgruppe  ''Theoretische Vielteilchenphysik''\\
         Universit\"at Rostock, D-18051 Rostock, Germany}
\author{V.N. Pervushin}
\address{Bogoliubov  Laboratory of Theoretical Physics,\\
        Joint Institute for Nuclear Research, 141980, Dubna, Russia}

\maketitle

\vspace{1cm}

\noindent MPG--VT--UR 67/95\\
\noindent December 1995

\begin{abstract}

Generalizing the Bogoliubov model of a weakly non-ideal Bose gas to massless
$\lambda\varphi^4$ theory we show that spontaneous breaking of symmetry occurs
due to condensation in a coherent vacuum
and leads to a vacuum energy density which is lower than that obtained by
Coleman and Weinberg using the one-loop effective potential method.
We discuss the alternative of a squeezed condensate and find that
for the massless $\lambda \varphi^4$ theory spontaneous symmetry
breaking to a squeezed vacuum does not occur.

\vspace{5mm}
\noindent
PACS number(s):  11.10.Gh, 11.15.Tk, 11.30.Qc
\end{abstract}

\newpage

There is little doubt  that the QCD vacuum has a highly
nontrivial structure characterized mainly by
the presence of a quark and a gluon condensate \cite{Shifman}.
These condensates, although not directly accessible by experiment, have a big
influence on many hadronic features.
Whereas the quark condensate as the consequence of
spontaneous chiral symmetry breaking of the fermions
has been the subject of many investigations \cite{hatsuda},
gluon condensation which seems to be a consequence of spontaneous gauge
symmetry breaking is still poorly understood.
In order to investigate the spontaneous breaking of a
bosonic symmetry, it is sensible first to consider
massless $\lambda\varphi^4$ theory as the simplest interacting
bosonic field theory which shares renormalisability and scale invariance
with QCD.

In 1973 Coleman and Weinberg \cite{CW} calculated the one loop effective
action of $\lambda\varphi^4$ and showed that the self-interaction of the
field may lead to spontaneous symmetry breaking and to the appearance of
a coherent condensate field $<\varphi>~\neq 0$. Since their approach was
limited to small $\lambda$ the meaning of the condensate field stayed unclear.

In 1982 Fr\"{o}hlich \cite{Frohlich} proved the triviality of
$\lambda\varphi^4$ by showing that for any finite bare coupling different from
zero the groundstate is unstable.

Castorina and Consoli \cite{Castorina} pointed out in 1983 that a new symmetric
phase ($<\varphi>~=0$, but $<\varphi^2>~\neq 0$) might exist with energy below
the trivial perturbative vacuum which is analogous to the case of superfluid
$^4 He$.
They use the formalism by Cornwall, Jackiw and Tomboulis \cite{Gaussian}
 to calculate the effective action for the composite operator $\varphi^2$.

Using the Gaussian effective potential method \cite{Gaussian}
Stevenson \cite{Stevenson} found that {\it massive} $\lambda\varphi^4$ theory
can have a metastable groundstate with spontaneous symmetry breaking
 and a coherent condensate
for an infinitesimally small negative bare coupling $\lambda$ (``precarious''
$\lambda\varphi^4$), which had escaped the earlier triviality proofs.
Two years later Stevenson and Tarrach \cite{Tarrach}
 found another, infinitesimally positive
bare coupling which might lead to a nontrivial theory with spontaneous
symmetry breaking, often called
``autonomous'' $\lambda\varphi^4$.
Whereas for the negative value of the bare coupling
there can be made a connection to perturbation theory, the positive one is
intrinsically nonperturbative, since it requires renormalisation of
the coherent zero momentum field. In both cases, the bare mass is
infinitesimally small, but not zero.
In 1992 Pedersen et al. \cite{Pedersen} have shown by construction the
existence of nontrivial {\it massless}  $\lambda\varphi^4$ theory in four
dimensions using the conformal invariance property.

Recent lattice calculations for massless $\lambda\varphi^4$,
proposed by Consoli and Stevenson \cite{ConStev} and performed by
Agodi et al. \cite{Agodi}, indicate that a coherent zero momentum condensate
indeed leads to spontaneous symmetry breaking.
The nonzero momentum modes, however, are massive but noninteracting modes.
Therefore, massless $\lambda\varphi^4$ with a coherent condensate seems to
be a trivial theory with a spontaneously broken vacuum.
We show that in massless  $\lambda\varphi^4$ theory a simpler approach can be
given to find the solution for a coherent vacuum. Furthermore, we  also
discuss the possibility of a squeezed vacuum  as an interesting alternative.

In this work we generalize the Bogoliubov model \cite{NN} of a weakly nonideal
Bose gas, invented to describe the low energy spectrum of superfluid Helium,
to massless $\lambda\varphi^4$ field theory.
Bogoliubov used the freedom due to the divergence of the Bose distribution
function to occupy the zero-momentum state macroscopically
by replacing the annihilation and creation operators $a_0$ and $a^+_0$ by the
very large number $\sqrt{N_0}$.
This corresponds to a coherent condensate of single zero momentum bosons and
leads to a spectrum of quasiparticle excitations which is that of Landau sound
for low momenta seen in experiment.

It is important to note that the Boguliubov result is also obtained
if one puts the weaker condition $a_0^2\sim a_0^{+2}\sim a_0^+a_0\sim N_0$
corresponding to a squeezed vacuum filled by zero-mode
quasiparticles which can be interpreted as pairs of bosons.
A vacuum which obeys $<\varphi_0>~=0$
but $<\varphi_0^2>~ >0$
is of great interest for investigations of the gluon condensate
in QCD \cite{Celenza,Pervush} where the
condensation of gluon pairs instead of single gluons would guarantee the
colour neutrality of a homogeneous vacuum.
In this letter we investigate both alternatives, tha coherent as well as the
squeezed vacuum of massless $\lambda\varphi^4$ theory.

Massless $\lambda\varphi^4$ theory is given by the Lagrangian density
\begin{equation} \label{ld}
{\cal L}(x) = \partial_\mu \varphi (x) \partial^\mu \varphi (x)
- \frac{\lambda}{4 !} \varphi ^4 (x)~.
\end{equation}
The corresponding Hamilton operator is
\begin{equation}
H[\varphi,\pi]  = \int  {\rm d} {\bf x} ~\left\{\frac{1}{2} \pi^2 ({\bf
x}) + \frac{1}{2}
(\partial_i\varphi({\bf x}) )^2 + \frac{\lambda}{4 !} \varphi^4 ({\bf
x})\right\}~,
\end{equation}
with the Schr\"odinger field operators $\varphi({\bf x}), \pi({\bf x})$
satisfying
 the canonical commutation relations
\begin{equation}
[\pi({\bf x}),\varphi({\bf x}')]= - i \delta({\bf x}-{\bf x}')~.
\end{equation}

In the momentum representation we have the Hamiltonian operator
\begin{equation}
H[\varphi,\pi]=
\frac{1}{2} \sum_{p}
\left[  \pi_p \pi_{-p}  + p^2 \varphi_p \varphi_{-p} \right]
+ {\lambda\over 4!V} \sum_{p_1 p_2 p_3 p_4}
\varphi_{p_1}\varphi_{p_2}  \varphi_{p_3}\varphi_{p_4}~
\delta_{p_1+p_2+p_3+p_4,0} ~,
\end{equation}
and the commutation relations $[\pi_p,\varphi_{p'}]= - i\delta_{p,-p'} ~$.
Furthermore, we have to introduce normal ordering
with respect to the creation and annihilation operators
$a_p, a_p^+$ defined by
\begin{eqnarray}
\label{phipi}
\varphi_p =\sqrt{\frac{1}{2  \tilde{\omega}(p)}} (a_p + a^+_{-p})\ , ~~
\pi_p = i \sqrt{\frac{\tilde{\omega}(p)}{2 }} (-a_p + a^+_{-p})     ~,
\end{eqnarray}
with an arbitrary function $\tilde{\omega}(p)$.
The operators $a_p, a_p^+$ satisfy the commutation relations
$[a_p, a_{p'}^+]= \delta_{p, p'}~$
The corresponding vacuum $|0> $ is defined by $a_p |0> ~= 0$,
 and the Fock space is given as
$\{\,|\,\Phi >\} = \{\,|\,0 >~~,~~  a^+_p |0 > = |\,p >,~~
  a^+_{p_1} a^+_{p_2} \,|\, 0 > = \,|\,{p_1 p_2} >, \cdots~\}$.
To each special choice of   $\tilde{\omega}(p)$ corresponds a certain
Fock space representation.

The particular choice
\begin{equation}
\label{omp}
\tilde{\omega}(p)=|p|~.
\end{equation}
diagonalizes the noninteracting ($\lambda$-independent) part of the
Hamiltonian, thus introducing a representation of free particles, which
are usually the basis states of a perturbative calculation for small values
of the coupling $\lambda$.
Note that for the dispersion relation (\ref{omp}) a singularity in
Eq. (\ref{phipi}) occurs for $p=0$ (the infrared singularity of massless
theories), so that the corresponding $a_0$ is not defined.
The zero mode $\varphi_0$ is not an oscillator-like variable
and has then to be treated separately.

In analogy to the Bogoliubov model we start with the
 representation (\ref{omp}) corresponding to free particles $a_p$ for $p\neq
0$.
Normal ordering of an operator $A$ with respect to the Fock space
representation of free particles will be denoted by the symbol $:A:$~.
The corresponding Hamiltonian is then written as
\begin{eqnarray}
\label{hf1}
:H[\varphi,\pi]:&=&
\frac{1}{2} \sum_{p\neq 0}
\left[ : \pi_p \pi_{-p} : + p^2 :\varphi_p \varphi_{-p} :\right]
+ {\lambda\over 4!V} \sum_{p_1 p_2 p_3 p_4 \neq 0} \delta_{P,0}
:\varphi_{p_1}\varphi_{p_2}  \varphi_{p_3}\varphi_{p_4} :
\nonumber\\
& & +\frac{1}{2}\pi_0^2
+ 6 {\lambda\over 4!V}\sum_{p\neq 0} :\varphi_p \varphi_{-p}: \varphi_0^2
+ {\lambda\over 4!V} \varphi_0^4~,
\end{eqnarray}
$V$ being the volume.
The transition to other Fock spaces can be given by unitary transformations.
In particular, we consider the coherent condensate of the zero-momentum
mode as well as the squeezed condensate.

The coherent condensate is
constructed using the unitary operator
\begin{equation}
U_{\rm coh} (\varphi_0,\pi_0) = \exp (i \alpha \pi_0).
\end{equation}
This unitary transformation shifts the field $\varphi_0$ to $\varphi_0+
\alpha$.
The coherent treatment of the Hamiltonian (\ref{hf1}) is obtained by replacing
the operator $\varphi_0$ by the macroscopically large c-number $\alpha$
characterizing the condensate.
Then we have $<\varphi_0^2>~=<\varphi_0>^2~=\alpha^2$ and
$<\varphi_0^4>~=\alpha^4$.

The construction of the squeezed condensate is achieved by applying the
unitary operator
\begin{equation}
\label{ub}
U_{\rm sq} (\varphi_0,\pi_0) =
\exp \left (i\frac{f_0}{2} (\pi_0 \varphi_{0} + \varphi_0 \pi_{0})
\right )~.
\end{equation}
Applying the unitary transformation (\ref{ub}), we can define
the new field operator $\varphi^{\rm sq}_0$  and its momentum $\pi^{\rm sq}_0$
by means of
\begin{eqnarray}
\label{bt}
\varphi^{\rm sq}_0 = U_{\rm sq} ^{-1}\varphi_0 U_{\rm sq} =
{\rm e}^{-f_0} \varphi_0~,\nonumber\\
\\
\pi^{\rm sq}_0 = U_{\rm sq} ^{-1}\pi_0 U_{\rm sq} = {\rm e}^{f_0} \pi_0
\nonumber~,
\end{eqnarray}
which satisfy the same algebra of commutation relations
as the initial ones.

Analogous to the Bogoliubov model we now  squeeze the zero-momentum mode
by the macroscopic parameter
$f_0$ so that
\begin{equation}
<0|(\pi^{\rm sq}_0)^2|0> ~\longrightarrow 0~, ~~~
<0|(\varphi^{\rm sq}_0)^2|0> ~\longrightarrow V ~C ~,
\end{equation}
where $C$ is positive.
In order to have similar notations for comparison with the coherent case
we write $\alpha^2=V~C$. In difference to the coherent case,
we have here $<0| (\varphi^{\rm sq}_0)^4 |0>~= 3~ (V~C)^2$

After shifting or squeezing the zero-momentum mode, the single particle part
of the Hamiltonian is
not diagonal any more in the free particles $a_p$ -- just as in the
Bogoliubov model -- and we have to move to a
representation of the nonzero components of the fields in terms of
quasiparticle operators $b_p$:
\begin{eqnarray}
\varphi_p =\sqrt{\frac{1}{2  \omega_B (p)}} (b_p + b^+_{-p})\ , ~~~
\pi_p = i \sqrt{\frac{\omega_B (p)}{2 }} (-b_p + b^+_{-p})~,
\ \ \ \ p\neq 0 ~.
\end{eqnarray}
For the time being, the dispersion relation $\omega_B (p)$ will be
considered as arbitrary and will be determined below by diagonalization
of the quasiparticle excitations.
The old and new operators $a_p$ and $b_p$ are related via the
 unitary Bogoliubov transformation
\begin{equation}
b_p =a_p \cosh f_p  + a^+_{-p} \sinh f_p ~,
\end{equation}
with the squeezing parameters
\begin{equation}
f_p={1\over 2}\ln {\omega_B (p)\over |p|}~.
\end{equation}
Changing from the original Fock space $|\Phi>$ to the new Fock space
$|\Phi_B>$ defined by the quasiparticle operators $b_p$, another normal
ordering procedure is necessary:
We have to reorder the Hamiltonian (\ref{hf1}) with respect to the new
vacuum $|0_B >~$ so that the operator $A$ is represented by the normal ordered
operator $::A::$ with respect to the new Fock space $|\Phi_B>$.
Reordering of the quadratic term gives
\begin{eqnarray}
\label{bcx}
: \varphi_{p_1} \varphi_{p_2} :
&=&::\varphi_{p_1} \varphi_{p_2}:: +~ \tilde{C}_{12}~,\nonumber\\
:\pi_{p_1} \pi_{p_2}: &=& ::\pi_{p_1} \pi_{p_2}:: + \tilde{C}_{12}^{\pi}~,
\end{eqnarray}
with
 \begin{eqnarray}
\label{bfi}
\tilde{C}_{12}&\equiv &
<0_{B} \,|\,\varphi_{p_1}\varphi_{p_2} \,|\,0_{B}>-
<0 \,|\,\varphi_{p_1}\varphi_{p_2} \,|\,0>=
{1\over 2}\delta_{p_1,-p_2}\left({1\over \omega_B (p_1)}-{1\over |p_1|}
\right)~,
\nonumber\\
\tilde{C}_{12}^{\pi}&\equiv &
<0_{B} \,|\,\pi_{p_1}\pi_{p_2} \,|\,0_{B}>-
<0 \,|\,\pi_{p_1}\pi_{p_2} \,|\,0>=
{1\over 2}\delta_{p_1,-p_2}\left(\omega_B (p_1)- |p_1|\right)~.
\end{eqnarray}
The quartic term is reordered according to
 \begin{eqnarray}
:\varphi_{p_1}\varphi_{p_2}\varphi_{p_3}\varphi_{p_4}:
& =&
::\varphi_{p_1} \varphi_{p_2} \varphi_{p_3} \varphi_{p_4}::
 +:: \varphi_{p_1} \varphi_{p_2}::
\tilde{C}_{34}  +(5\,\, {\rm permutations}) \nonumber\\
& &+ \tilde{C}_{12}\tilde{C}_{34} +
(2 \,\,{\rm permutations})~.
\end{eqnarray}
As result we get the  reordered  Hamiltonian (\ref{hf1}) in  the form
\begin{eqnarray}
\label{hfb}
:H[\varphi,\pi]: =
E_0
+ ::H^{(2)}[\varphi,\pi]::
+ ::H^{(4)}[\varphi]::~,\nonumber
\end{eqnarray}
where a vacuum energy  shift $E_0=<0_B|H|0_B>$ occurs,
which depends on the structure of the zero momentum mode as expressed by the
values for the coherent condensate
\begin{eqnarray}
\label{h0}
E_0^{\rm coh}&=&{\lambda V\over 4!}
C^2
  +\frac{1}{2} \sum_{p\neq 0}
\left[\tilde{C}_{p,-p}^{\pi}+\left(p^2+ {\lambda\over 2  }C
      \right)\tilde{C}_{p,-p}\right]\nonumber\\
& &
+{\lambda\over 4!V}\left(\sum_{p \neq 0}
\tilde{C}_{p,-p}^2
+ 3 \sum_{p_1\neq p_2 \neq 0}
\tilde{C}_{p_1,-p_1}\tilde{C}_{p_2,-p_2}\right)~,
\end{eqnarray}
and the squeezed condensate
\begin{eqnarray}
\label{h1}
E_0^{\rm sq}=E_0^{\rm coh}+ {\lambda V\over 12} \left[C^2 + \frac{1}{V^2}
\sum_{p \neq 0} \tilde{C}_{p,-p}^2 \right]~.
\end{eqnarray}
Furthermore, we find
\begin{eqnarray}
\label{h2}
 ::{H}^{(2)}[\varphi,\pi]::
&=&\frac{1}{2}\sum_{p\neq 0}\left\{::\pi_{p}\pi_{-p}::  +
\left[p^2 +  {\lambda\over 2}C
+  {\lambda\over 2 V} \sum_{p' \neq 0} \tilde{C}_{p',-p'}\right]
::\varphi_{p} \varphi_{-p}::\right\} ~,\nonumber\\
::{H}^{(4)}[\varphi]:: &=& {\lambda\over 4!V}\sum_{p_1,p_2,p_3,p_4 \neq 0}
\delta_{P,0}
::\varphi_{p_1}\varphi_{p_2}\varphi_{p_3}\varphi_{p_4} :: ~.
\end{eqnarray}
As a consequence, there occurs in $::{H}^{(2)}[\varphi,\pi]::$ a quadratic term
in $\varphi$  which defines the Bogoliubov mass,
\begin{equation}
\label{mass}
m_B ^2\equiv  {\lambda\over 2}\left[{C }
+ {1\over V} \sum_{p\neq 0} \tilde{C}_{p,-p}\right]~.
\end{equation}

A diagonal form in terms of the $b_p, b^+_p$ ($p\neq 0$) is achieved if we put
\begin{equation}
\omega_B (p)= \sqrt{p^2+m_B ^2}~.
\end{equation}
The squeezing parameters $f_p$ are then given by
\begin{equation}
f_p ={1\over 4} \ln{p^2 + m_B ^2\over p^2}~~.
\end{equation}
We see that for large momenta $p\rightarrow\infty$ there is no squeezing
($f_p\rightarrow 0$),
whereas for $p\rightarrow 0$ the squeezing becomes very large
($f_p\rightarrow \infty$).

With this determination of  $\omega_B (p)$,  Eq. (\ref{mass}) becomes
\begin{equation}
\label{mass2}
m_B ^2= {\lambda\over 2} \left[ C + {1 \over {2V}} \sum_{p \neq 0}
\left( \frac{1}{\sqrt{{p}^2 + m_B ^2}} - \frac{1}{|p|}\right) \right]~,
\end{equation}
which has the form of a gap equation for $m_B $.
The vacuum energy density  $\epsilon \equiv E_0/V$ for both cases becomes
\begin{eqnarray}
\label{eps0}
\epsilon ^{\rm coh} &=&
{1\over 4 V}\sum_{p \neq 0}\left[\left(\sqrt{{p}^2 + m_B ^2} - |p|\right)
  -\left(\frac{p^2}{\sqrt{{p}^2 + m_B ^2}} - |p|\right)\right] \nonumber\\
& &+ \frac{\lambda}{8}
\left[ C +{1\over {2 V}}\sum_{p\neq 0}\left({1\over\sqrt{{p}^2 + m_B ^2}}
   -{1\over |p|}\right)\right]^2 \nonumber\\
& &  - \frac{\lambda}{12}\left[ C^2
+{1\over 4V^2}\sum_{p\neq 0}\left({1\over\sqrt{{p}^2 + m_B ^2}}
   -{1\over |p|}\right)^2\right] ~,\\
\epsilon ^{\rm sq} &=& \epsilon ^{\rm coh}
+ \frac{\lambda}{12} \left[C^2+{1\over 4V^2}\sum_{p\neq 0}
\left({1\over\sqrt{{p}^2 + m_B ^2}} -{1\over |p|}\right)^2\right]~.
\end{eqnarray}
Both the gap equation and the expressions for $\epsilon $ are logarithmically
ultraviolet divergent and have to be regularized by introducing a momentum
cutoff $\Lambda$.
In the expression for $\epsilon $ we use the gap equation to simplify the
second term. We obtain in the limit $V \rightarrow \infty$
\begin{eqnarray}
\label{mbreg}
m_B^2 &=& \frac{\lambda}{2} \left[ C - \frac{m_B^2}{8 \pi^2}
\ln \left(2 \frac{\Lambda}{m_B}\right) + \frac{1}{16 \pi^2} m_B^2
+ {\cal O}\left(\frac{m_B^2}{\Lambda^2}\right) \right]~,\\
\label{eps1}
\epsilon ^{\rm coh} &=&
{m_B ^4\over 32\pi^2}\left[\ln\left(2 {\Lambda \over m_B }
\right)-{1\over 8} \right]+{m_B ^4\over 2\lambda}
-{\lambda\over 12} C^2 +{\cal O}\left(\frac{m_B^2}{\Lambda^2}\right) ~,\\
\epsilon ^{\rm sq} &=& \epsilon ^{\rm coh} + \frac{\lambda}{12} C^2 ~.
\label{esq}
\end{eqnarray}
This can be made finite by using the renormalized coupling
constant $\lambda$ which results from the regularized gap equation
(\ref{mbreg}),
\begin{eqnarray}
\label{lamb}
\lambda^{-1}  &=& -\frac{1}{16 \pi^2} \left[ - \frac{8 \pi^2}{m_B^2} C +
\ln \left(2 \frac{\Lambda}{m_B}\right) - \frac{1}{2}  \right]
+ {\cal O}\left(\frac{m_B^2}{\Lambda^2}\right) ~,
\end{eqnarray}
for any $m_B$ and leaving the density $C$ at some  fixed
$\Lambda$ as a parameter.
For $C(\Lambda)$ we make the ansatz
\begin{equation}
\label{c}
C(\Lambda)\equiv A {m_B^2\over 8\pi^2}
\ln\left(2\frac{\Lambda}{m_B}\right)+ C_0~,
\end{equation}
where $A$ and $C_0$ are the parameters of the ansatz, so that
\begin{equation}
\label{lambda}
\lambda^{-1}(\Lambda)=
-{1\over 16\pi^2}\left[(1-A)\ln\left(2{\Lambda\over m_B }\right)
-\frac{1}{2}\right]+\frac{C_0}{2m_B ^2}~.
\end{equation}
Inserting these expressions into (\ref{eps1}) we find that
the logarithmic ultraviolet divergences in $\epsilon^{\rm coh}$ cancel
for the two cases $A=0$ and $A=3$, corresponding to an infinitesimally
small negative and an infinitesimally small positive bare coupling constant,
respectively.

In the case $A=0$, $C$ remains finite and the logarithmic singularity in the
vacuum energy density is cancelled by that of the bare coupling (\ref{lamb}).
In this case the energy density $\epsilon_-^{\rm coh}$ for infinitesimally
small negative bare coupling constant becomes
\begin{equation}
\label{eco-}
\epsilon_-^{\rm coh}={3\over 8}{m_B ^4\over 32\pi^2}+{m_B ^2\over 4} C_0~.
\end{equation}
Since $C=C_0>0$, the minimum is at $m_B =0$ and there is no spontaneous
symmetry breaking.

The other case $A=3$ is obtained assuming a compensating divergency also
in $C$.
In this case we have an infinitesimally small positive coupling and the
energy density $\epsilon_+^{\rm coh}$ becomes
\begin{equation}
\epsilon_+^{\rm coh} ={9\over 8}{m_B ^4\over 32\pi^2}+{m_B ^2\over 8}C_0~.
\end{equation}
The leading term of Eq. (\ref{c}) is positive, therefore $C_0$ can be negative.
For $C_0<0$ this expression has the form of a mexican hat as a function of
$m_B $.
Minimizing $\epsilon_+^{\rm coh} $ with respect $m_B ^2$ we find
\begin{equation}
C_0=-{9\over 16\pi^2}m_B ^2
\end{equation}
with the value
\begin{equation}
\label{eco+}
\epsilon_+^{\rm coh} =-{9\over 8}{m_B ^4\over 32\pi^2}
\end{equation}
at the minimum.
Here a nonzero value of $m_B $ is therefore favoured for $C_0<0$.
For both solutions of the coherent condensate, our values of $\lambda$
from Eq. (\ref{lambda}) correspond exactly to those obtained by Stevenson
\cite{Stevenson}.
Our value (\ref{eco+}) of $\epsilon_+^{\rm coh}$ lies below the value
$-(1/4)m_B ^4/(32\pi^2)$ of Consoli and Stevenson \cite{ConStev},
who used the Coleman-Weinberg formula for
the one-loop effective potential and made it manifestly finite by performing
a renormalization of the coherent field with a logarithmically divergent
factor similar to our ansatz (\ref{c}).

For a sqeezed condensate of the zero-momentum mode, only the solution for
$A=0$ remains,
\begin{equation}
\label{epsq}
\epsilon^{\rm sq} = \epsilon_-^{\rm coh}~.
\end{equation}
This result is obtained because
the second term in Eq. (\ref{esq}) does not contribute for $A=0$ since $C$
is finite and $\lambda \longrightarrow 0 $.
Since (\ref{epsq}) does not lead to spontaneous symmetry breaking as discussed
after (\ref{eco-}), the  squeezed condensate does not occur in massless
$\lambda \varphi^4$ theory.

In conclusion, we found that for the massless $\lambda \varphi^4$  theory
no squeezed condensate occurs but a coherent condensate leading to
a quasiparticle spectrum with finite mass $m_B$  .
This ''Bogoliubov'' mass $m_B$ follows in a selfconsistent way from a gap
equation containing the mean square $<\varphi_0^2>~=V~C$ of the zero-momentum
fluctuations as an ingredient.

The gap equation is used as a condition for the possible bare couplings which
renormalize the ultraviolet divergent energy density.
In the case of negative bare coupling constant the resulting finite energy
density in the coherent and in the squeezed case coincide but do not lead to
spontaneous symmetry breaking.

For positive bare coupling, spontaneous symmetry breaking occurs.
For massless scalar $\lambda \varphi^4$ theory, this leads
only to a coherent and not to a squeezed condensate.
Our value (\ref{eco+}) for the vacuum density lies below that of
Consoli and Stevenson \cite{ConStev} obtained from the one-loop effective
potential of Coleman and Weinberg.
For a coloured $\lambda \varphi^4$ model and for QCD the investigation of the
squeezed condensate is an interesting subject of subsequent work.

The  models for a condensate presented here provide a transparent
renormalisation procedure related to the parameter $C$ for the condensate.
We have given a simple derivation of an effective Hamiltonian in massless
$\lambda \varphi^4$ theory which allows for the description of spontaneous
symmetry breaking due to the occurrence of a vacuum condensate.

\section*{Acknowledgement}
We thank M. Teper for suggestions and P. Petrow for a careful reading of the
manuscript.
One of us (V.N.P.) acknowledges the financial support provided
by the Max-Planck-Gesellschaft and the hospitality of the MPG Arbeitsgruppe
''Theoretische Vielteilchenphysik'' at the Rostock University.

\end{document}